\documentclass[aps,pra,reprint,superscriptaddress,showpacs]{revtex4-1}
\usepackage{graphicx,amsmath,amssymb,ulem}
\usepackage{verbatim}
\usepackage{ragged2e}
\usepackage{braket}
\usepackage{amsmath}

\begin{document}
\title{Dark-state and loss-induced phenomena in the quantum-optical regime of \ensuremath{\Lambda}-type three-level systems}


\author{H.~Rose}
\affiliation{Department of Physics and CeOPP, University of Paderborn, Warburger Stra\ss{}e 100, D-33098 Paderborn, Germany\looseness=-1}
\
\author{D.~V.~Popolitova}
\affiliation{Faculty of Physics, Moscow State University, Leninskie Gory, 1, Moscow, 119991 Russia}
\
\author{O.~V.~Tikhonova}
\affiliation{Faculty of Physics, Moscow State University, Leninskie Gory, 1, Moscow, 119991 Russia}
\
\author{T.~Meier}
\affiliation{Department of Physics and CeOPP, University of Paderborn, Warburger Stra\ss{}e 100, D-33098 Paderborn, Germany\looseness=-1}
\ 
\author{P.~R.~Sharapova}
\affiliation{Department of Physics and CeOPP, University of Paderborn, Warburger Stra\ss{}e 100, D-33098 Paderborn, Germany\looseness=-1}


\begin{abstract}

The interaction of matter with quantum light leads to phenomena which cannot be explained by semiclassical approaches. Of particular interest are states with broad photon number distributions which allow processes with high-order Fock states. Here, we analyze a Jaynes-Cummings-type model with three electronic levels which is excited by quantum light. As quantum light we consider coherent and squeezed states. In our simulations we include several loss mechanisms, namely, dephasing, cavity, and radiative losses which are relevant in real systems. We demonstrate that losses allow one to control the population of electronic levels and may induce coherent population trapping, as well as lead to a redistribution of the photon statistics among the quantum fields and even to a transfer of the photon statistics from one field to another. Moreover,  we introduce and analyze a novel quantity, the quantum polarization, and demonstrate its fundamental difference compared to the classical polarization. Using the quantum polarization and the third level population, we investigate electromagnetically induced transparency in the presence of quantum light and show its special features for the case of squeezed light. Finally, quantum correlations between fields are studied and analyzed in the presence of different types of losses.

\end{abstract}

\maketitle

\section{INTRODUCTION}

Three-level systems (3LS) are a brilliant concept in optics as they are frequently used for the approximate description of atomic systems as well as semiconductors and semiconductor nanostructures in situations where predominantly two transitions can be excited \cite{Scully,HaugKoch,MeierThomasKoch}.
Here, we consider a \ensuremath{\Lambda} system \cite{popolitova2019,PhysRevA.54.1586,Wang,PRX}
which has been utilized for both, the study of optical excitations with classical light \cite{PhysRevA.94.063411,PhysicsLettersA.80.140}, and for effects that are exclusive observed for the quantum-optical regime \cite{ncomms12303}. \ensuremath{\Lambda} systems can be realized in different ways: from optically pumped semiconductor quantum wells enclosed in optical microcavities \cite{PRL108,klettke2013} to charged semiconductor quantum dots \cite{nphys1054,PRL118} and doped semiconductors in strong magnetic fields \cite{PRL95}.

Prominent applications that were established by the use of \ensuremath{\Lambda}-type 3LS are coherent population trapping (CPT) \cite{CPT1,CPT2,Scully} and electromagnetically induced transparency (EIT) \cite{Scully,EIT1,EIT2,EIT3}, which both describe the formation of dark states \cite{PhysRevA.99.053829}. While these phenomena are well studied for classical light, the quantum-optical regime is poorly investigated and needs a careful analysis.

Quantum light has several advantages compared to classical light. The quantum-mechanical description of light as photons allows for different photon statistics, leading to a class of states with unique properties. Important examples are single photon states \cite{Michler,McKeever,Dariquie,Maunz}, which are a central requirement for quantum-information processing \cite{PhysRevA.79.032303} and quantum cryptography \cite{PhysRevLett.89.187901}, as well as squeezed states \cite{Slusher,Phys.Rev.Lett.57,Phys.Rev.A.38}, that allow to reduce the noise level below the shot noise limit \cite{Slusher} and were successfully applied to reduce quantum noise in gravitational wave detectors \cite{Barsotti_2018}. The use of quantum light also leads to novel phenomena, such as entanglement between light and matter \cite{ent}, where a description of the matter with a 3LS is at the origin of quantum memories \cite{qmemnat}, quantum repeaters \cite{QRepeater}, optical storage \cite{OStorage}, and all-optical neural networks \cite{ONN}. Therefore, fundamental properties of the interaction between quantized light and 3LS are of great interest.

In this article, we investigate the interaction between 3LS and different kinds of quantum light. Extending a previous investigation \cite{popolitova2019} we demonstrate dark state phenomena such as  EIT and CPT in the quantum-optical regime in the presence of losses. In contrast to CPT, EIT was not yet demonstrated within a quantum-optical model taking into account photon statistics of quantum fields. In this work, we demonstrate it for coherent and squeezed vacuum states. Since for some quantum fields the classical polarization vanishes,  we introduce a new measure to characterize a system interacting with quantum light - a quantum polarization. The quantum polarization describes the specific response of the system to quantized light and contains an information about light-matter correlations. We use the quantum polarization as well as the time-averaged population for the demonstration of EIT spectra. The quantum polarization is compared with the classical polarization and its benefits are discussed. Finally, we demonstrate and investigate the influence of losses on quantum-correlations between fields.

\section{THEORETICAL MODEL}\label{sec:thmod}

\begin{figure}[ht]
	\centering
	\includegraphics[width=0.25\textwidth]{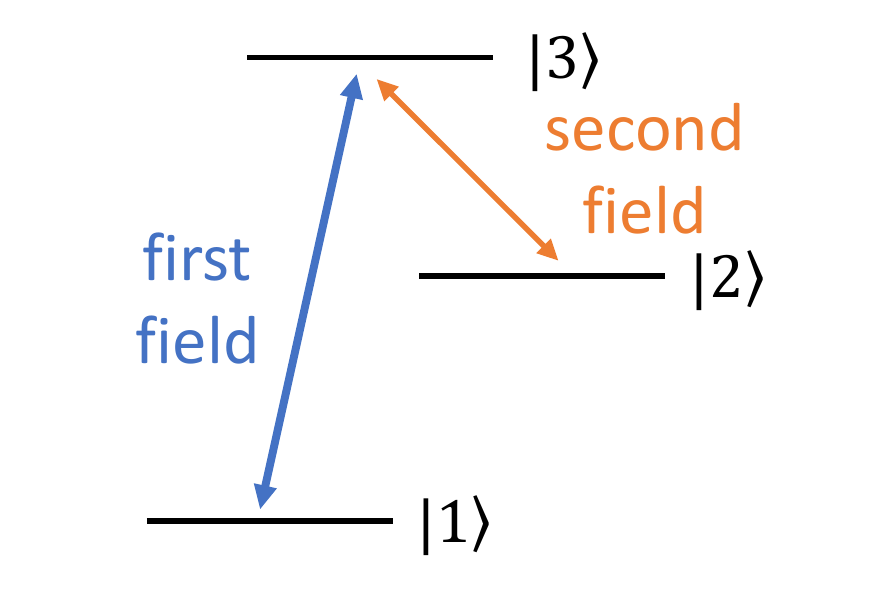}
	\caption{(color online) Illustration  of the considered three-level system.}
	\label{fig:3LS}
\end{figure}

In this work we consider a Jaynes-Cummings-type model:  the interaction between a 3LS and two quantum fields. The electronic states of the 3LS are denoted with $\ket{1}$ for the ground state and $\ket{2}$ and $\ket{3}$ for the excited states and we consider that initially only state $\ket{1}$ is fully populated.
The transition between $\ket{1}$ and $\ket{2}$ is dipole-forbidden, while the other transitions are allowed. We consider two single-mode quantum fields: the first field excites the transition between levels $\ket{1}$ and $\ket{3}$, while the second field  excites the transition between levels $\ket{2}$ and $\ket{3}$. This setup is known as the \ensuremath{\Lambda} scheme and illustrated in Fig.~\ref{fig:3LS}. 

The full Hamiltonian describing the model can be written as the sum of the Hamiltonians corresponding to matter, light fields, and their interaction:
\begin{align}
\hat{H} = \hat{H}_{\mathrm{3LS}} + \hat{H}_{\mathrm{L,P}} + \hat{H}_{\mathrm{L,C}} + \hat{H}_{I,P,1} + \hat{H}_{I,C,2},
\label{eq:Hamilton}
\end{align}
where the index $L$ denotes the field Hamiltonians, $I$ - the interaction Hamiltonians, $P$ corresponds to the first field, and $C$ - to the second field. 
With the transition operator $\hat{\sigma}_{ij}=\ket{i}\bra{j}$ and bosonic creation (annihilation) operators $\hat{a}^{\dagger}_i$ ($\hat{a}_i$) for the mode $i$, the Hamiltonians read
\begin{align}
\hat{H}_{\mathrm{3LS}} &= \sum_{n=1}^{3} E_n \hat{\sigma}_{nn},\\
\hat{H}_{\mathrm{L,i}} &= \hbar \omega_i \bigg( \hat{a}^{\dagger}_i \hat{a}_i +\frac{1}{2} \bigg),\\
\hat{H}_{\mathrm{I,i,j}} &= -\frac{\mu_{j3}\varepsilon_{0i}}{\sqrt{2}}\Big(\hat{a}^{\dagger}_i + \hat{a}_i\Big)\Big(\hat{\sigma}_{j3} + \hat{\sigma}_{3j}\Big),
\label{eq:SubHamilton}
\end{align}
where $E_n$ are the energies of the electronic levels, $\mu_{ij}$ is the dipole matrix element for the $\ket{i} \rightarrow\ket{j}$ transition and $\varepsilon_{0i}=\sqrt{\frac{4\pi\hbar\omega_i}{V}}$ is the constant field amplitude where $V$ denotes the interaction volume \cite{Zapyantsev2018}. Applying the rotating-wave approximation (RWA) leads to the omission of the terms $\hat{a}^{\dagger}_i\hat{\sigma}_{3j}$ and $\hat{a}_i \hat{\sigma}_{j3}$ in the light-matter interaction.
The von Neumann equation for the density matrix $\hat{\rho}$ including losses reads:
\begin{align}
\partial_t \hat{\rho} = \frac{1}{i\hbar}[\hat{H}, \hat{\rho}]_{-} + \sum_{\hat{L}} \hat{\mathcal{L}}_{\hat{L}}(\hat{\rho}),
\label{eq:Neumann}
\end{align}
where the Lindblad term $\hat{\mathcal{L}}_{\hat{L}}(\hat{\rho})$ describes losses by the operator $\hat{L}$ \cite{Lindblad1976}. In the Schrödinger picture, the Lindblad term reads:
\begin{align}
\hat{\mathcal{L}}_{\hat{L}}(\hat{\rho}) = \frac{1}{2} \bigg( 2 \hat{L} \hat{\rho} \hat{L}^{\dagger} - \hat{L}^{\dagger} \hat{L} \hat{\rho} - \hat{\rho}\hat{L}^{\dagger} \hat{L}\bigg).
\label{eq:Lindblad}
\end{align}
We take into account three loss mechanisms: cavity losses, radiative losses, and dephasing losses with parameters $\kappa$, $r$, and $\gamma$, respectively. The cavity losses arise from the finite lifetime of a photon in the cavity and are modeled with the operator $\hat{L} = \sqrt{\kappa} \hat{a_i}$. Furthermore, the recombination of electrons from excited states to the ground state without emission of a photon leads to radiative losses which are modeled with $\hat{L} = \sqrt{r_{i,j}} \ket{i}\bra{j}$ for $E_{j} > E_{i}$. These losses are also applied to the dipole forbidden transition and take into account higher-order processes. Moreover, material polarizations can decay due to various processes, e.g., scattering with phonons or Coulomb scattering. This is described by dephasing losses which are modeled with $\hat{L}=\sqrt{\gamma_{i,j}}(\ket{i}\bra{i} - \ket{j}\bra{j})$ for $i>j$ \cite{Schneebeli2010}. The dephasing losses will be applied only to the non-diagonal elements since they decay more rapid than the diagonal ones.

The density matrix is composed of three subsystems, namely the electronic levels and two light modes, and can be written in the general form:
\begin{align}
\hat{\rho} = \sum_{\substack{n=1 \\ n'=1}}^3\sum_{\substack{k=k'=0 \\ m=m'=0}}^\infty \rho_{\substack{n,k,m \\ n',k',m'}}\ket{n,k,m}\bra{n',k',m'},
\label{eq:DM1}
\end{align}
where $n$ denotes the electronic state which can either be $1$, $2$, or $3$ and $k$ and $m$ denote Fock states of the first and the second field, respectively. In the initial moment of time, the electronic and fields subsystems are given by

\begin{align}
\ket{M} &= \sum_{n=1}^{3} c^M_n \ket{n},\\
\ket{P} &= \sum_{k=0}^\infty c^P_k \ket{k},\\
\ket{C} &= \sum_{m=0}^\infty c^C_m \ket{m},
\end{align}
where $c^M_n$ are the probability amplitudes for an electron to be in the state $\ket{n}$ and $ c^P_k$ and $ c^C_m$ are the probability amplitudes to find the first field in the Fock state $\ket{k}$ and the second field in the Fock state $\ket{m}$, respectively.

Therefore, the statistics of the quantum fields are incorporated in the initial condition for Eq.~(\ref{eq:Neumann}). In this paper, we consider coherent and squeezed vacuum states at the initial moment of time. A coherent state $\ket{\alpha}$ gives the quantum-mechanical description of classical laser light and is characterized by the following probability amplitudes $c_k$ and the mean photon number $\braket{\hat{n}}=\braket{\hat{a}_i^{\dagger}\hat{a}_i}$:
\begin{align}
c_k &= e^{-\frac{|\alpha|^2}{2}} \frac{\alpha^k}{\sqrt{k!}},\\
\braket{\hat{n}} &= |\alpha|^2.
\label{eq:coherent}
\end{align}
The probability amplitudes and the mean photon number of a squeezed state $\ket{\xi}$ read
\begin{align}
&c_{2k} = (-1)^k \sqrt{\frac{2\beta}{1 + \beta^2}} \frac{\sqrt{(2k)!}}{2^k k!} \bigg( \frac{1-\beta^2}{1+\beta^2} \bigg)^k,\\
&c_{2k+1} = 0,\\
&\braket{\hat{n}} = \frac{1}{4} \bigg( \beta - \frac{1}{\beta} \bigg)^2,
\label{eq:sq}
\end{align}
where $\xi=e^{-|\beta|}$ is the squeezing parameter. Squeezed vacuum states are characterized by zero population of all odd states, however, this situation can be changed due to different processes, e.g. interaction with matter or losses. To denote the mean photon number of the first and the second field, we use $N_P$ and $N_C$  respectively. In the case of $\braket{\hat{n}}=N_P=N_C$,  we will use only $\braket{\hat{n}}$.

The population of the electronic states $\mathcal{O}_n$ and the photon distribution of the first $W_k$ and the second $\tilde{W}_m$ fields can be extracted by tracing out the density matrix over the other variables 
\begin{align}
\mathcal{O}_n(t) &= \sum_{k=0}^{\infty} \sum_{m=0}^{\infty}  \rho_{\substack{n,k,m \\ n,k,m}},\\
W_k(t) &= \sum_{n=1}^{3} \sum_{m=0}^{\infty}  \rho_{\substack{n,k,m \\ n,k,m}},\label{eq:phstat}\\
\tilde{W}_m(t) &= \sum_{n=1}^{3} \sum_{k=0}^{\infty}  \rho_{\substack{n,k,m \\ n,k,m}}.
\end{align}
The polarization response for classical fields is proportional to the dipole moment, averaged over the time-dependent wave function \cite{HaugKoch}. This would result in our case in the following expression for a classical polarization response to the first field:
\begin{align}
P^C_{31}(t) &= \sum_{k=0}^{\infty} \sum_{m=0}^{\infty}  \rho_{\substack{3,k,m \\ 1,k,m}}.
\label{eq:cp}
\end{align}
 
During the interaction of matter with quantum fields, the light and the matter subsystems become correlated and the averaged dipole moment can be zero for certain initial states of light. This leads to a vanishing classical polarization response because an excitation of electronic levels is connected with a corresponding change in the field statistics.

Thus, one should introduce a new measure, the quantum polarization response, where both the dipole operator and the field operator are taken into account. 
This quantum polarization for a chosen electronic transition can be obtained from the density matrix by tracing out one of the fields. For example, for the transition  between levels $\ket{1}$ and $\ket{3}$ initiated by the first field, the quantum polarization reads
\begin{align}
\hat{P}_{31}(t) &= \sum_{m=0}^{\infty}  \rho_{\substack{3,k,m \\ 1,k',m}}\ket{k}\bra{k'}.
\end{align}
Since we suppose only single-photon absorption and emission processes, only the elements which fulfill $k = k' - 1$ are considered. Thus, the macroscopic quantum polarization can be written as
\begin{align}
P^Q_{31}(t) &= \sum_{k=0}^{\infty} \sum_{m=0}^{\infty}  \rho_{\substack{3,k,m \\ 1,k+1,m}}.
\label{eq:qp}
\end{align}
In a numerical simulation, the sum is truncated by introducing the maximum photon numbers $k_{\mathrm{max}}$ and $m_{\mathrm{max}}$ that can be chosen differently for both fields and strongly depend on the initial field statistics. Since the number of elements in the density matrix increases with the fourth power of the maximum photon number, the numerical evaluation is a demanding task for high photon numbers. However, the problem can be simplified using the following substitution:
\begin{align}
\rho_{\substack{n,k,m \\ n',k',m'}} = p_{\substack{n,k,m \\ n',k',m'}} \exp\bigg(\frac{1}{i \hbar}(E_{n, k, m} - E_{n', k', m'})t\bigg),
\label{eq:decomposition}
\end{align}
where $E_{n,k,m}=E_n + E_k + E_m$ is the total energy composed of electronic state and photon energies. Subsequently, detunings for the excitations can be introduced as $\Delta_P = E_3 - E_1 - (E_{k+1} - E_k)$ for the first and $\Delta_C = E_3 - E_2 - (E_{m+1} - E_m)$ for the second field. After applying the RWA, Eq.~(\ref{eq:Neumann}) is transformed to 
\onecolumngrid
\begin{align}
\begin{split}
\partial_t p_{\substack{n,k,m \\ n',k',m'}}(t)=\Omega_1\frac{i}{\sqrt{2}}\bigg(&p_{\substack{n+2,k-1,m \\ n',k',m'}}(t)e^{i\Delta_P t}\sqrt{k}+p_{\substack{n-2,k+1,m \\ n',k',m'}}(t)e^{-i\Delta_P t}\sqrt{k+1}\\
&-p_{\substack{n,k,m \\ n'+2,k'-1,m'}}(t)e^{-i\Delta_P t}\sqrt{k'}-p_{\substack{n,k,m \\ n'-2,k'+1,m'}}(t)e^{i\Delta_P t}\sqrt{k'+1}\bigg)\\
+\Omega_2\frac{i}{\sqrt{2}}\bigg(&p_{\substack{n+1,k,m-1 \\ n',k',m'}}(t)e^{i\Delta_C t}(1-\delta_{n,1})\sqrt{m}+p_{\substack{n-1,k,m+1 \\ n',k',m'}}(t)e^{-i\Delta_C t}(1-\delta_{n,2})\sqrt{m+1}\\
&-p_{\substack{n,k,m \\ n'+1,k',m'-1}}(t)e^{-i\Delta_C t}(1-\delta_{n',1})\sqrt{m'}-p_{\substack{n,k,m \\ n'-1,k',m'+1}}(t)e^{i\Delta_C t}(1-\delta_{n',2})\sqrt{m'+1}\bigg)\\
+\frac{\kappa}{2}\bigg[&2p_{\substack{n,k+1,m \\ n',k'+1,m'}}(t)\sqrt{k+1}\sqrt{k'+1}+2p_{\substack{n,k,m+1 \\ n',k',m'+1}}(t)\sqrt{m+1}\sqrt{m'+1}\\
&-p_{\substack{n,k,m \\ n',k',m'}}(t)\big(k+k'+m+m'\big)\bigg]-\gamma_{n,n'} p_{\substack{n,k,m \\ n',k',m'}}(t) (1 - \delta_{n,n'})\\
&+\frac{1}{2} \sum_{i=1}^{2} r_{i,3} \bigg[ p_{\substack{3,k,m \\ 3,k',m'}}(t) 2 \delta_{n,i} \delta_{n,n'} - p_{\substack{n,k,m \\ n',k',m'}}(t) (\delta_{3,n}+\delta_{3,n'})\bigg]\\
&+\frac{r_{1,2}}{2}\bigg[ 2 p_{\substack{2,k,m \\ 2,k',m'}}(t) \delta_{1,n}\delta_{n,n'}-p_{\substack{n,k,m \\ n',k',m'}}(t)(\delta_{2,n} + \delta_{2,n'} ) \bigg].
\label{eq:Motion}
\end{split}
\end{align}
\twocolumngrid
This set of differential equations allows to identify the elements that are relevant for the dynamics. Without losses and detunings, Eq.~(\ref{eq:Motion}) can be solved analytically \cite{popolitova2019}.

In the following, we use times and frequencies in units of Rabi periods $\frac{1}{\Omega_1}$ and Rabi frequencies $\Omega_1$, respectively, and use $\Omega_2=\Omega_1$ for all simulations, where $\Omega_i = \frac{1}{\hbar}\mu_{i3}\varepsilon_{0i}$ is the Rabi frequency for the respective transition. For a coherent state with a mean photon number of $\braket{\hat{n}}\le 100$ or a squeezed state with a mean photon number of $\braket{\hat{n}}\le 10$, $k_{\mathrm{max}}=200$ is chosen, while for a squeezed state with a mean photon number of $\braket{\hat{n}}\le 100$, $k_{\mathrm{max}}=1200$ is chosen.

\section{RESULTS AND DISCUSSION}

In this Section, we present and discuss our theoretical considerations and the results of our numerical simulations. Firstly, it is shown that the population dynamics and the photon statistics can be manipulated by loss mechanisms. After that, we investigate the EIT effect within the quantum-optical regime and compare the classical and the quantum polarizations. Lastly, the impact of losses on quantum-optical correlations between the two quantum fields is studied.
Henceforth, the initial condition for the electronic system is chosen as $c_1^M=1$ and $c_2^M=c_3^M=0$.

\subsection{Manipulation of Population Dynamics with Losses}\label{sec:pop}

\begin{figure}[ht]
	\centering
	\includegraphics[width=0.5\textwidth]{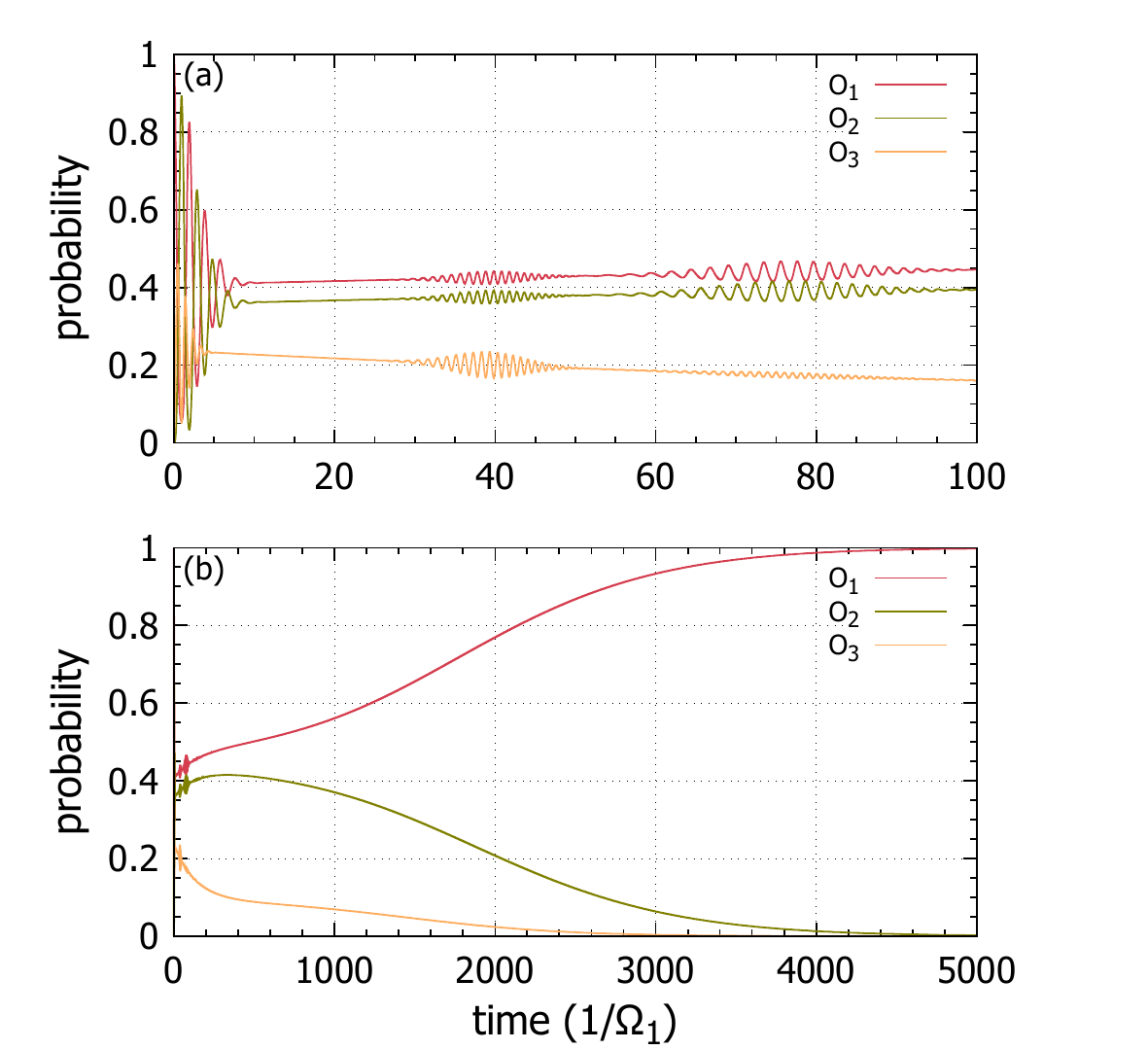}
	\caption{(color online) Time evolution of the electronic level populations with dephasing, radiative, and cavity losses. The loss parameters are set to $\kappa = 0.001\Omega_1$, $r_{1,3}=r_{2,3}=0.01\Omega_1$, $r_{1,2}=0.002\Omega_1$, $\gamma_{3,1}=\gamma_{3,2}=0.01\Omega_1$, and $\gamma_{2,1}=0.002\Omega_1$. Two exciting coherent states with $\braket{\hat{n}}=10$ are considered. The dynamics is shown up to (a) $t\Omega_1=100$ and (b) $t\Omega_2=5000$.}
	\label{fig:Fig2}
\end{figure}

In this section, we show that losses can influence the dynamics of observables in unexpected and nontrivial ways and may induce interesting effects such as CPT.  We start our analysis by considering the population dynamics for the excitation by two coherent states with mean photon numbers of $\braket{\hat{n}}=10$ in the presence of weak losses, see Fig.~\ref{fig:Fig2}. The loss parameters are chosen as $\kappa = 0.001\Omega_1$, $r_{1,3}=r_{2,3}=0.01\Omega_1$, $r_{1,2}=0.002\Omega_1$, $\gamma_{3,1}=\gamma_{3,2}=0.01\Omega_1$, and $\gamma_{2,1}=0.002\Omega_1$.

We note that in the time range shown in Fig.~\ref{fig:Fig2}(a), the well-known behavior of collapses and revivals is seen in the level populations induced by coherent fields. However, due to the presence of losses, all level populations are finally transferred to the ground state, i.e., $\mathcal{O}_1=1$ and $\mathcal{O}_2=\mathcal{O}_3=0$, as can be seen on the long time scale shown in Fig.~\ref{fig:Fig2}(b). This is caused by the finite photon flux in the quantum-optical regime. All excitations from the finite photon flux are lost over time due to cavity losses and thus the system reverts to the ground state again. In contrast, continuous wave classical fields have a steady photon flux by which steady states different from the ground state can be formed.

\begin{figure}[ht]
	\centering
	\includegraphics[width=0.5\textwidth]{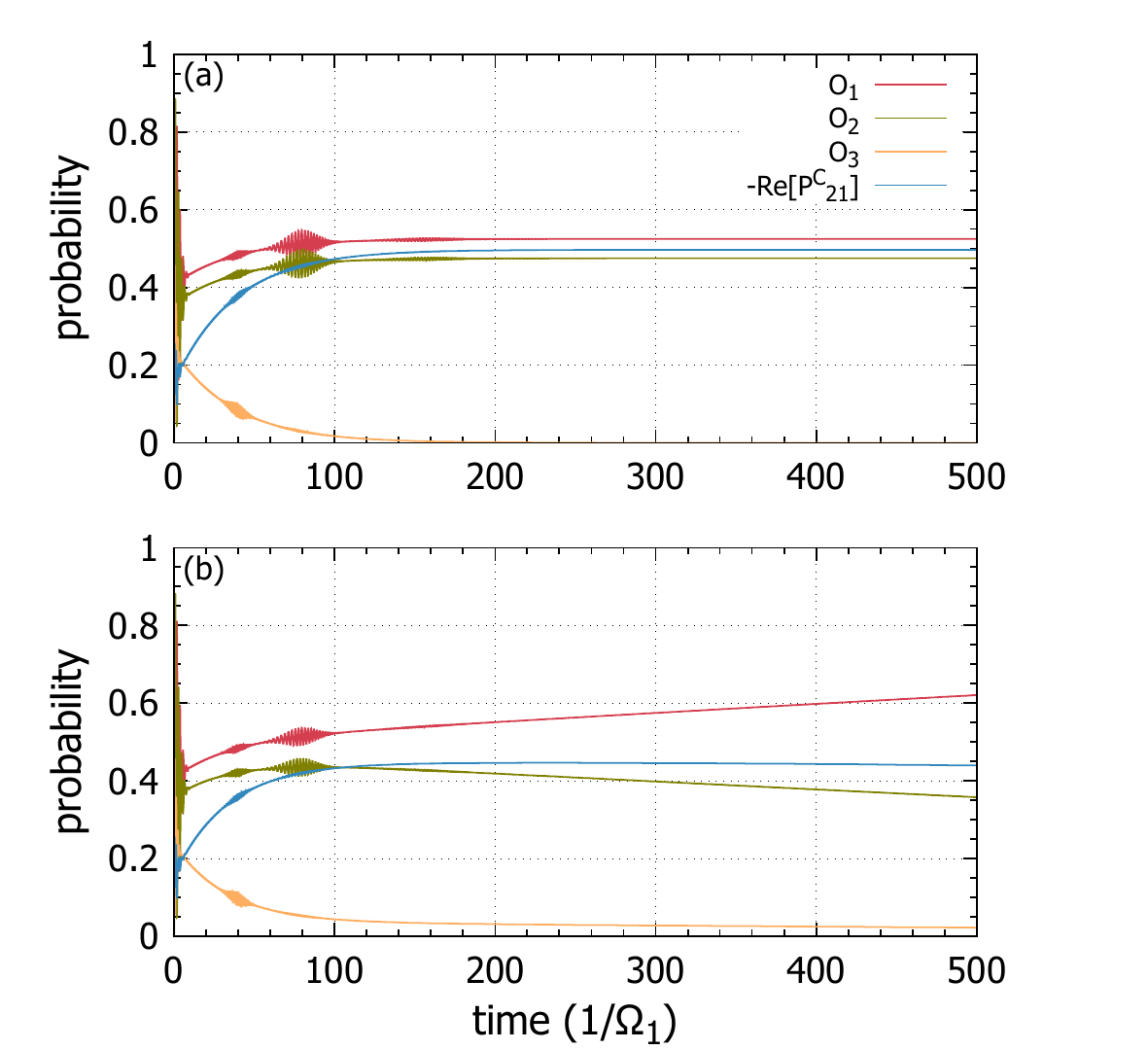}
	\caption{(color online) Time evolution of the population and the coherence $P^C_{21}$ with losses for two coherent states with mean photon numbers of $\braket{\hat{n}}=10$. Cavity and dephasing losses are neglected, while the parameters for the radiative losses are set to $r_{1,3}=r_{2,3}=0.05\Omega_1$ and in (a) $r_{1,2}=0$ whereas in (b) $r_{1,2}=0.01\Omega_1$.}
	\label{fig:Fig3}
\end{figure}

Contrary to a conventional treatment, the presence of losses enables the control of the population dynamics, which we will demonstrate below focusing on CPT. In atomic three-level systems with the  first and the second levels populated initially, CPT can be achieved by introducing an initial relative phase between the populated atomic states, the states of light, or the dipole matrix elements \cite{popolitova2019,Scully}. This is not possible when only the lowest electronic level is initially populated.
However, activated radiative losses $r_{1,3}$  and $r_{2,3}$ lead to a fast decay of the third level, whereby almost equal populations of the first and the second levels can be achieved, which is demonstrated in Fig.~\ref{fig:Fig3}(a). For this analysis, we assume high quality cavities that provide negligibly small photon losses, so that we can study the effects induced by radiative losses without taking the cavity losses into account. Therefore, we consider quasi steady-states, that are induced by radiative losses at a point in time where the photon losses do not have a visible impact.

This result is better understood by considering the reduced density matrix, traced over the states of light:
\begin{align}
\hat{\rho}_{\mathrm{red}} = \sum_{k,m}^\infty \bra{k,m}  \rho_{\substack{n,k',m' \\ n',k'',m''}}\ket{n,k',m'}\braket{n',k'',m''|k,m}.
\label{eq:reducedDM} 
\end{align}
We can identify the diagonal elements as populations and the non-diagonal elements or coherences as classical polarizations, i.e., $\braket{i|\hat{\rho}_{\mathrm{red}}|i} = \mathcal{O}_i$ and $\braket{i|\hat{\rho}_{\mathrm{red}}|j} = P^C_{ij}$, respectively. The classical polarization for the transition $\ket{i}\rightarrow\ket{j}$ is calculated with Eq.~(\ref{eq:cp}), by substituting indices $1$ and $3$ with $i$ and $j$. Note that considering $i,j\in\{1,2\}$ is sufficient in the steady state, since $\mathcal{O}_3=0$. Therefore, Fig.~\ref{fig:Fig3}(a) fully describes the reduced density matrix. An important quantity in this context is the Schmidt number $K$, which is given by:
\begin{align}
K=\frac{1}{\mathrm{Tr}[\hat{\rho}^2_{\mathrm{red}}]},
\label{eq:schmidtfull}
\end{align}
where $\mathrm{Tr}$ is the trace. In the case discussed above, the expression for the Schmidt number simplifies to:
\begin{align}
K = \frac{1}{\mathcal{O}^2_1 + \mathcal{O}^2_2 + 2|P^C_{21}|^2},
\label{eq:schmidt} 
\end{align}
which is found to be very close to one in the steady state and therefore the electronic subsystem is essentially in a pure state. Thus, we find a coherent superposition of levels $\ket{1}$ and $\ket{2}$. The classical polarization carries information about the relative phase of these levels. Denoting the absolute phases of levels $\ket{1}$ and $\ket{2}$ with $\phi_1$ and $\phi_2$, respectively, we can rewrite the classical polarization for a pure steady state as:
\begin{align}
P^C_{21 st} = \sqrt{\mathcal{O}_1 \mathcal{O}_2} e^{i(\phi_2 - \phi_1)},
\end{align}
which equals to $P^C_{21}$ for $K=1$. Since $\mathrm{Im}[P^C_{21}] = 0$ and $\mathrm{Re}[P^C_{21}] < 0$ in the steady state, we can conclude that the electronic levels $\ket{1}$ and $\ket{2}$ have a relative phase of $\pi$ and CPT is realized.

Fig.~\ref{fig:Fig3}(b) shows a more realistic scenario, in which a finite $r_{1,2}$ is introduced. In this case, losses are applied to the dipole-forbidden transition also and therefore the CPT is finally destroyed. However, since the $r_{1,2}$ losses are small, we find a large time interval where CPT is still present.

\begin{figure}[ht]
	\centering
	\includegraphics[width=0.5\textwidth]{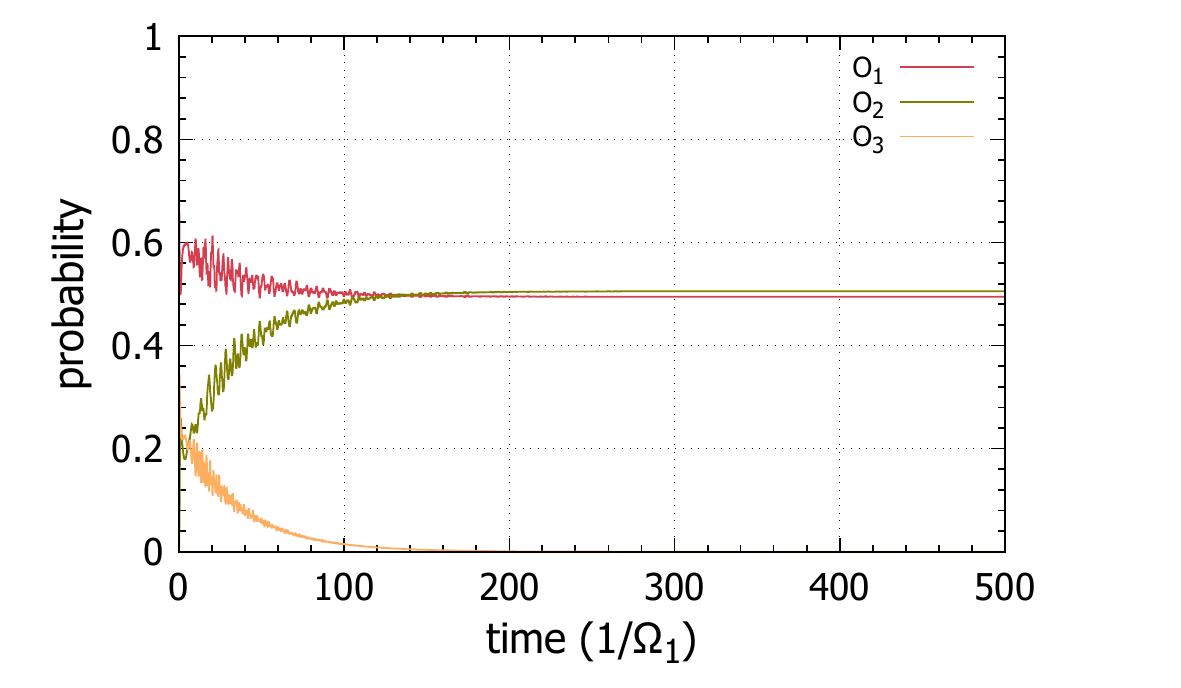}
	\caption{(color online) Time evolution of the population with losses for two squeezed vacuum states with a mean photon number of $N_P=10$ and $N_C=4$, respectively. Cavity and dephasing losses are neglected, while the parameters for the radiative losses are set to $r_{1,3}=r_{2,3}=0.05\Omega_1$ and $r_{1,2}=0\Omega_1$.}
	\label{fig:Fig4}
\end{figure}

For the case of two squeezed vacuum states, we can also achieve almost equal populations of the first and the second electronic levels in the steady state by applying losses, see Fig.~\ref{fig:Fig4}. However, due to complete vanishing of the classical polarization for the squeezed vacuum light $P^C_{21}=0$, in the steady state regime the Schmidt number is $K=2$. This means that the electronic subsystem is in a mixed state and there is no CPT. Thereby, in contrast to coherent fields, squeezed vacuum fields do not induce CPT but lead to the entanglement of electronic levels which were initially in a pure state.  

\subsection{Manipulation of Photon Statistics with Losses}

The controlled manipulation of the dynamics is not limited to the electronic level populations, but can also be applied to photon statistics. For this analysis, we again neglect cavity losses and consider quasi steady-states, as was mentioned in Sec.~\ref{sec:pop}.

\begin{figure}[ht]
	\centering
	\includegraphics[width=\columnwidth]{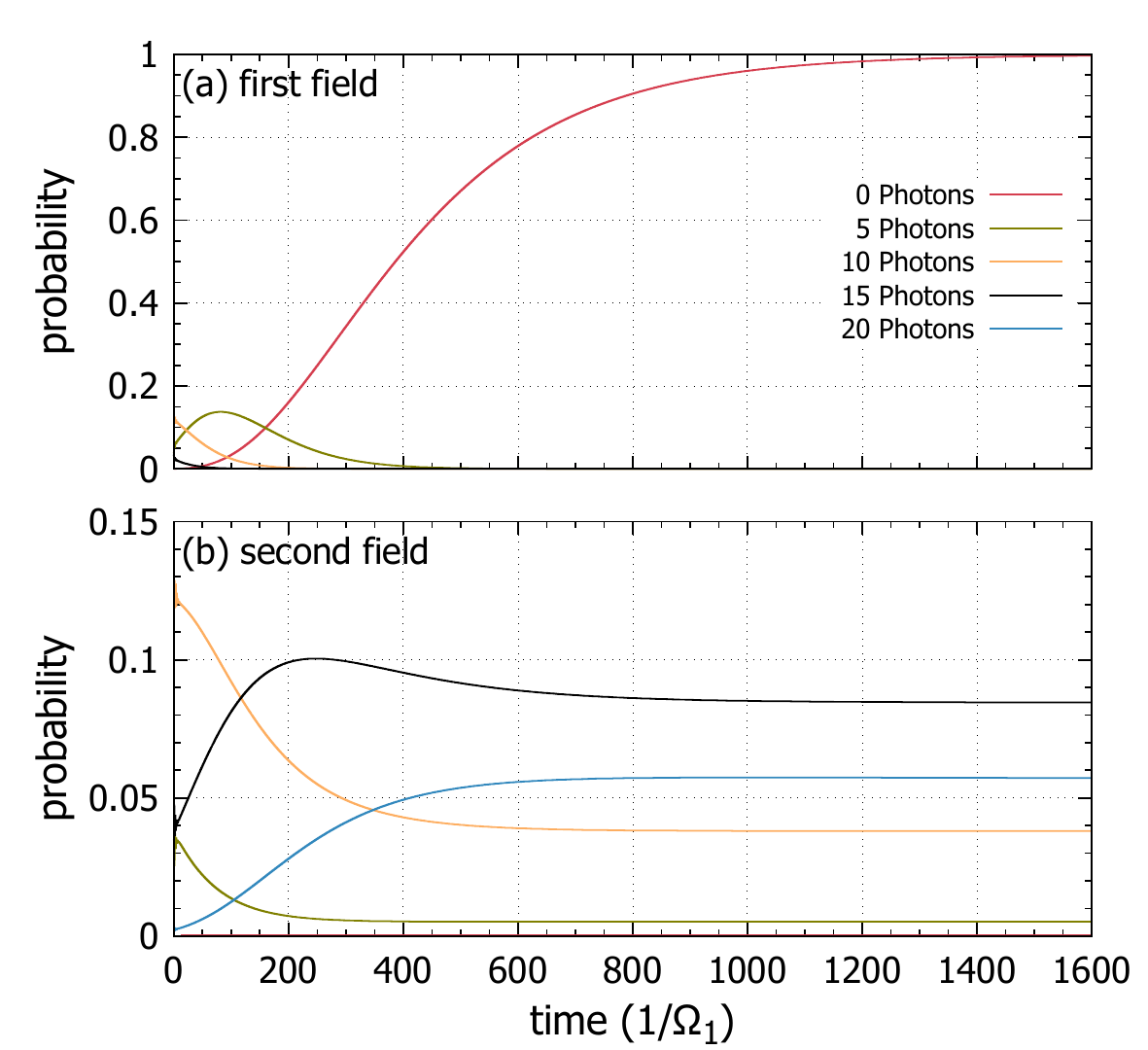}
	\caption{(color online) Dynamics of the photon statistics for  two incident coherent states with $\braket{\hat{n}}=10$ is presented for (a) the first field ($W_k$) and (b) the second field ($\tilde{W}_m$). Radiative losses are applied with $r_{1,3}=r_{2,3}=0.5 \Omega_1$ and $r_{1,2}=0.1\Omega_1$.}
	\label{fig:Fig5}
\end{figure}

Fig.~\ref{fig:Fig5} demonstrates the dynamics of $W_k$ and $\tilde{W}_m$ induced by two coherent fields  with mean photon numbers of  $\braket{\hat{n}}=10$ in the presence of radiative losses.
One can see that even though the photon statistics of the first field is reverted to the vacuum state, this is not the case for the photon statistics of the second field $\tilde{W}_m$ in the steady state. Rather, the photon statistics is redistributed.  The redistribution of the photon statistics is caused by higher-order processes described by $r_{1,2}$. Due to this loss mechanism, the transition from the second to the first level is possible without a change in the number of photons. Therefore, in the special case of a finite $r_{1,2}$, while all other losses are zero, the photons are transferred from the first to the second field.

\begin{figure}[ht]
	\centering
	\includegraphics[width=0.5\textwidth]{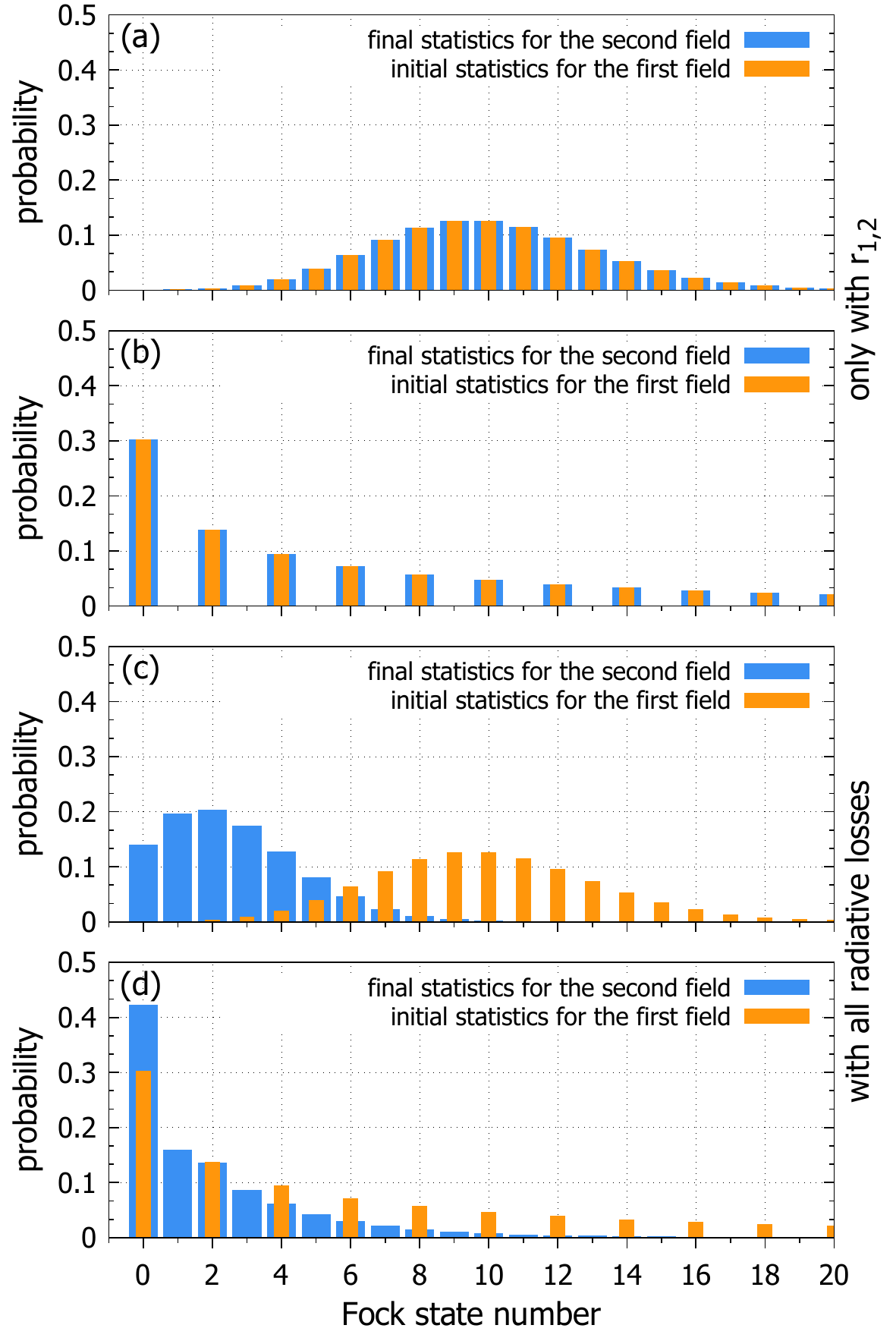}
	\caption{(color online) The initial photon statistics of the first field $W_k$ (orange bars) is shown together with the photon statistics of the second field $\tilde{W}_m$ at $t\Omega_1 = 2000$ (blue bars), where in (a) and (b) only  $r_{1,2}=0.5\Omega_1$ is considered, while in (c) and (d) $r_{1,3}=r_{2,3}=0.05\Omega_1$ and $r_{1,2}=0.01\Omega_1$ are chosen. The initial photon statistics of the first field is chosen as a coherent field in (a) and (c)  and as a squeezed field in  (b) and (d) with $\braket{\hat{n}}=10$, respectively. The second field initially is in the vacuum state.}
	\label{fig:Fig6}
\end{figure}

Starting from a vacuum state for the second field leads to a complete transfer of the photon statistics from the first to the second field, which is demonstrated in Figs.~\ref{fig:Fig6}(a) and \ref{fig:Fig6}(b). Such scheme gives the possibility to obtain information about an unknown input field  by measuring the properties of the second field. In a more realistic scenario, i.e. $r_{1,3}=r_{2,3}=0.05\Omega_1$ and $r_{1,2}=0.01\Omega_1$, this transfer is less ideal, i.e., the final photon statistics of the second field is not the same as the original photon statistics of the first field, see Figs.~\ref{fig:Fig6}(c) and \ref{fig:Fig6}(d). However, in all cases the first field is in the vacuum state in the end, which means that all photons from the first field are transferred completely.

\begin{figure}[ht]
	\centering
	\includegraphics[width=0.5\textwidth]{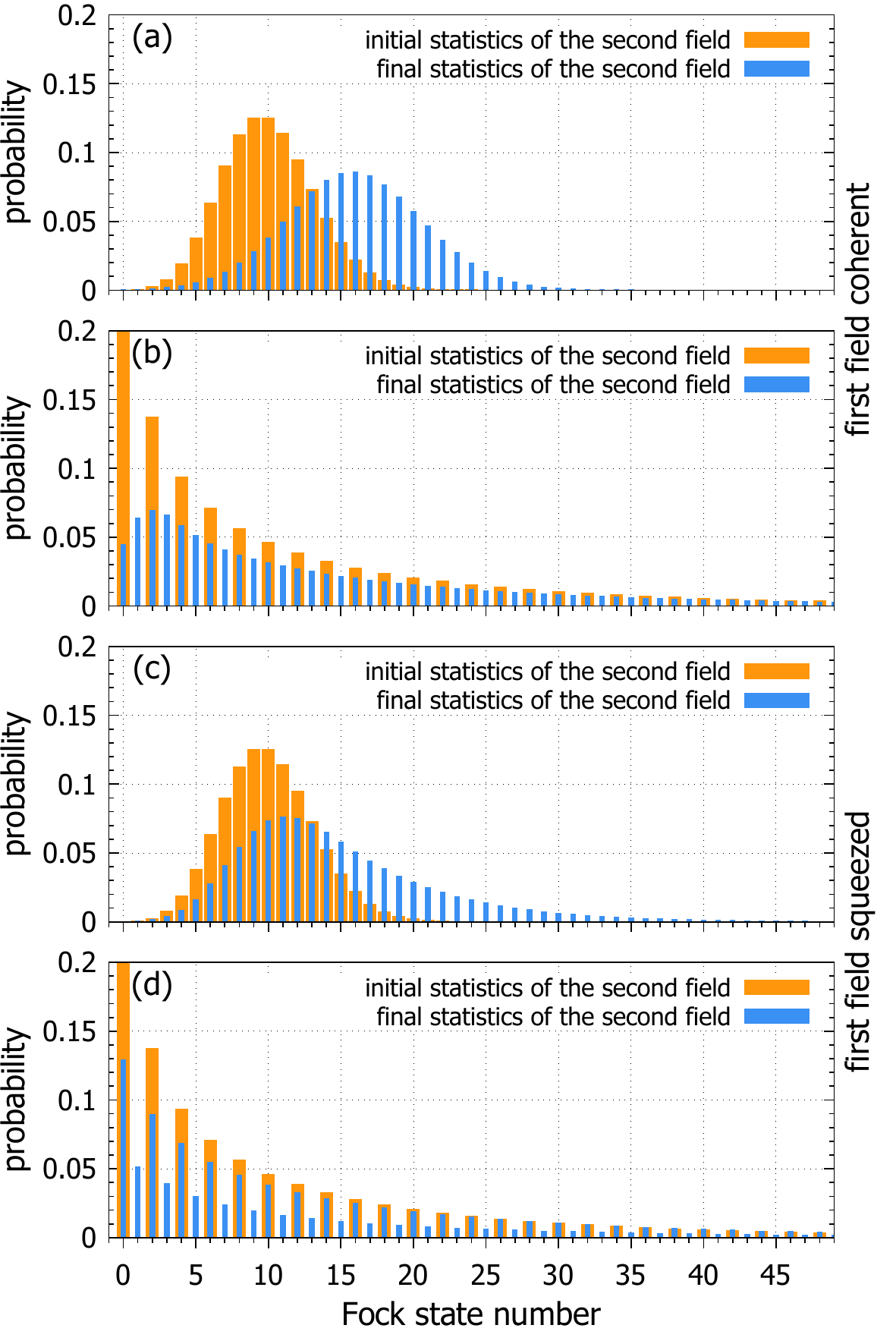}
	\caption{(color online) Photon statistics of the second field $\tilde{W}_m$ at the initial moment of time (orange bars) and for $t\Omega_1 = 2000$ (blue bars) when the steady state is formed. Different  combinations of the first and the second fields are considered in the initial moment of time: (a) two coherent states, (b) a coherent and a squeezed vacuum state, (c) a squeezed vacuum and a coherent state, and (d) two squeezed vacuum states. All initial fields have a mean photon number $\braket{\hat{n}}=10$.  $r_{1,3}=r_{2,3}=0.5\Omega_1$ and $r_{1,2}=0.1\Omega_1$ are chosen. }
	\label{fig:Fig7}
\end{figure}

If the second field has a special statistics in the initial moment of time, the final distribution of the second field reflects features of this statistics. Fig.~\ref{fig:Fig7} demonstrates the photon number probability distribution of the second field $\tilde{W}_m$ at the initial moment of time (orange bars) and at the moment of time when the steady state is formed (blue bars). Four different combinations of coherent and squeezed states of light with the mean photon number of $\braket{\hat{n}}=10$  are considered at the initial moment of time. One can see that features of the initial photon statistics of both, the first field $W_k(0)$ and the second field $\tilde{W}_m(0)$, strongly influence the steady-state distribution of the second field. For the case of two coherent states, see Fig.~\ref{fig:Fig7}(a), the resulting distribution is similar to a coherent state but has a higher mean photon number $\braket{\hat{n}}\approx 16$. In the case of two squeezed vacuum states, see Fig.~\ref{fig:Fig7}(d), the resulting distribution has a mean photon number of $\braket{\hat{n}}\approx 13.8$ and shows similarities to the squeezed vacuum state but has a non-zero probability of measuring odd photons due to activated radiative losses $r_{1,3}$ and $r_{2,3}$. The mean number of photons were directly calculated from the resulting photon statistics of the second field and show that coherent states lead to a more efficient transfer of photons in comparison to squeezed states.  In both cases, the final photon statistics cannot be described with the initial photon statistics of the first or the second field separately, but still shows a similar to them behavior. For the case of two different initial photon statistics, see Figs.~\ref{fig:Fig7}(b) and \ref{fig:Fig7}(c), we find almost the same mean photon numbers of $\braket{\hat{n}}\approx 14.9$ and $\braket{\hat{n}}\approx 14.8$, respectively, but completely different final statistics of the second field. The final state of the first field is the vacuum state in all cases.

Thereby, the second field may inherit properties of the photon statistics of the first field due to the higher-order processes described by $r_{1,2}$. This allows a redistribution of the photon statistics, whose quality depends on the other losses.

\subsection{Electromagnetically Induced Transparency}\label{sec:EIT}

EIT is an effect in which an otherwise opaque medium can be rendered transparent for a resonant probe field in the presence of a coupling field, which is well understood from a semiclassical theory \cite{EIT2} and was observed experimentally \cite{EIT1,EIT3}. This effect is based on destructive interference between the transitions from the ground state to two quasienergy states describing the upper level dressed by the strong coupling field. The destructive interference found at the resonance frequency is lifted for detuned excitations, which is why an EIT spectrum has a minimum at zero detuning and two peaks next to it \cite{Scully, EIT1, EIT2, EIT3}.

\begin{figure}[ht]
	\centering
	\includegraphics[width=0.45\textwidth]{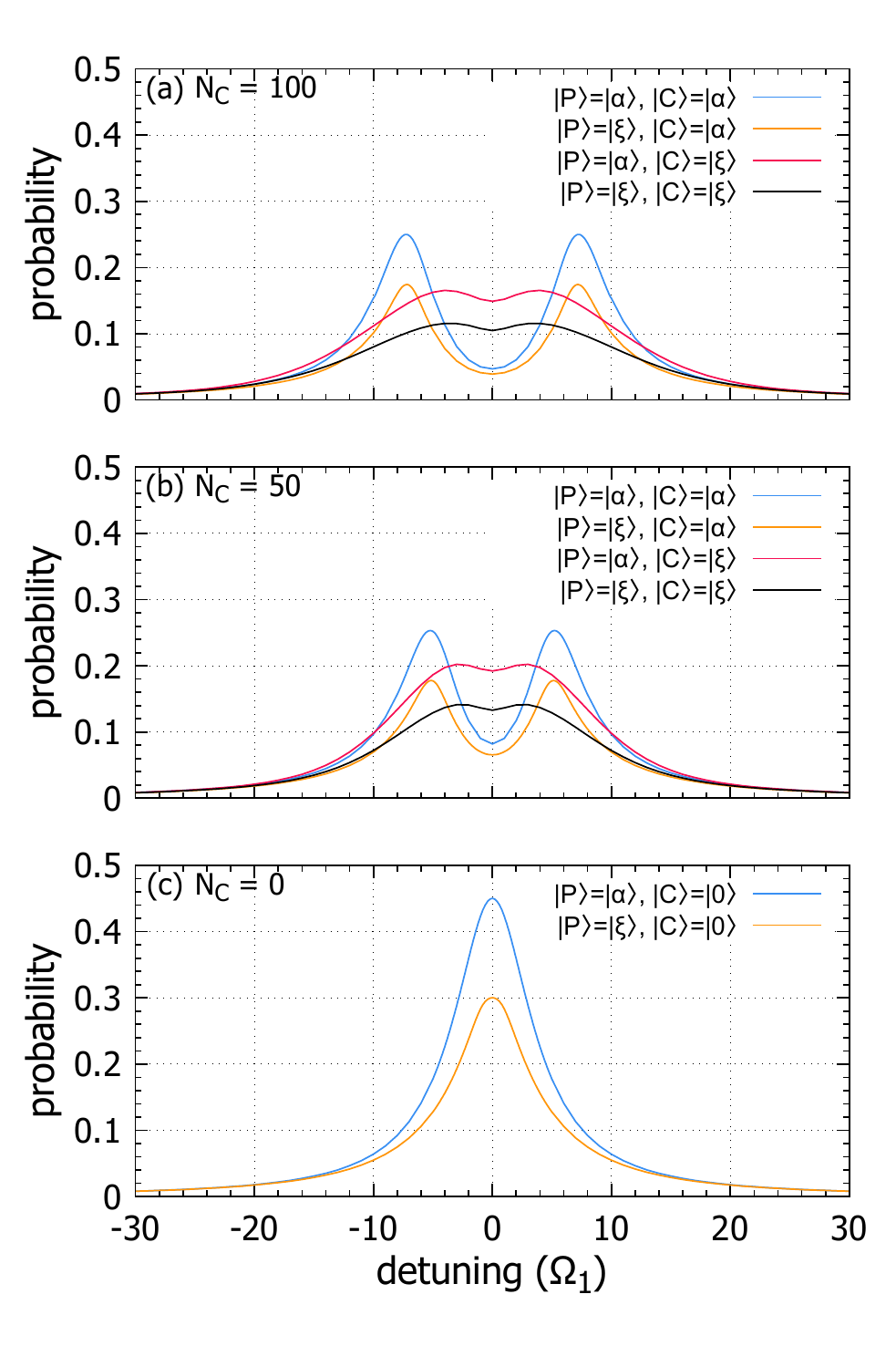}
	\caption{(color online) The time-averaged population of the third electronic level as a function of the detuning for different combinations of light states. Here, $N_P=10$ and (a) $N_C=100$, (b) $N_C=50$, and (c) $N_C=0$ is chosen and no losses are applied, i.e., $\kappa=r_{i,j}=\gamma_{ij}=0$.}
	\label{fig:EITPop}
\end{figure}

The use of quantum light leads to novel features in EIT which we demonstrate in this section. With the system shown in Fig.~$\ref{fig:3LS}$, we consider the first field as the probe and the second field as the coupling field.
Due to the chosen initial condition where only the first electronic level is filled, while the others are empty, the time-averaged population of the third level can be considered as a measure for absorption. Thus, the detuning-dependent time-averaged third level population yields an absorption spectrum. The data for the averaging was calculated from
$t\Omega_1=0$ to $t\Omega_1=100$ with $\Delta t \Omega_1= 0.01$. To simulate a strong coupling field, $N_P=10$ is chosen, whereas $N_C$ is chosen as $50$ and $100$, respectively. The calculated EIT spectra for different combinations of initial states of light are shown in  Fig.~\ref{fig:EITPop} with a decrease in the mean photon number of the coupling field from (a) to (c).

It is demonstrated that the spectra obtained by using two coherent states are similar to the semiclassical result but are not its direct reproduction due to a set of collapses and revivals in the dynamics of the electronic populations in the quantum-optical regime. The form of the absorption spectra is mainly determined by the photon statistics of the coupling field, whereas the probe field changes their magnitude. Figs.~\ref{fig:EITPop}(a) and \ref{fig:EITPop}(b) demonstrate that a squeezed state as a probe field results in less absorption across the detuning range, therefore leading to a more efficient EIT. In the quantum regime, properties of the probe field directly influence the absorption. To demonstrate this, we consider the vacuum state  as the coupling field, while the choice of $N_P=10$ for the coherent and the squeezed probe field remains unchanged, see Fig.~\ref{fig:EITPop}(c). In this case one can see that the probability of promoting electrons to the excited state is higher for a coherent state than for a squeezed state. This means that the absorption for a squeezed probe field is lower than for a coherent probe field with the same mean photon number. This effect is directly connected with the photon statistics. The excitation induced by squeezed light is less efficient due to a large contribution of the zeroth-Fock state which cannot initiate transitions between electronic levels. Thereby, due to the presence of the high-populated vacuum component, a certain amount of the electronic population cannot be promoted. This property of squeezed light remains in the EIT regime, therefore, squeezed states can be applied to reduce absorption.

The lineshapes of the EIT spectra can be explained from the point of view of the dressed states. The splitting of the quasienergy levels for squeezed coupling light is smaller in comparison to coherent states. This is the reason why the maxima in the case of a squeezed coupling field are very close to each other, leading to no pronounced minimum in the center, see Figs.~\ref{fig:EITPop}(a) and \ref{fig:EITPop}(b). But this splitting increases with the increase of the number of photons of the coupling field, which can be seen by comparing Fig.~\ref{fig:EITPop}(a) and Fig.~\ref{fig:EITPop}(b). This allows us to interpret the EIT process as a probing of the dressed light-matter state, which was created by the interaction of matter with the coupling field.

The EIT spectra can also be understood quantitatively by considering the quasienergy states. The position of the maxima induced by a coupling field with the initial photon statistics $c^C_m$ can be estimated by the following expression:
\begin{align}
\Omega_g = \pm \frac{\Omega_1}{\sqrt{2}} \sum_{m}^\infty |c^C_m|^2 \sqrt{m}.
\label{eq:splitting}
\end{align}
Evaluating this expression for a coherent state with mean photon numbers of $100$ and $50$ leads to $\Omega_g=\pm7.06\Omega_1$ and $\Omega_g=\pm4.99\Omega_1$, respectively. These values are in a good agreement with the positions of the maxima in the corresponding spectra, found at $\pm 7.2\Omega_1$ and $\pm 5.2\Omega_1$, respectively. Note that the values obtained using Eq.~(\ref{eq:splitting}) are an estimation, since Eq.~(\ref{eq:splitting}) does not take into account the fact that each Fock state forms its own quasienergy and furthermore, the probe field is not taken into account. 

A similar comparison for squeezed states leads to $\Omega_g=\pm5.58\Omega_1$ and $\Omega_g=\pm3.91\Omega_1$ for $N_C=100$ and $N_C=50$, respectively. One can see that these values are not in a good agreement with the peaks in the EIT spectra, which are found at around $\pm3.6\Omega_1$ and $\pm2.5\Omega_1$, respectively. This difference arises since the peaks in the EIT spectra are not clearly separated but overlap, which results in an overall shift of the maxima. Nevertheless, this estimation shows that the splitting for the case of squeezed vacuum is smaller than for coherent states.

\begin{figure*}[ht]
	\centering
	\includegraphics[width = 1.6\columnwidth]{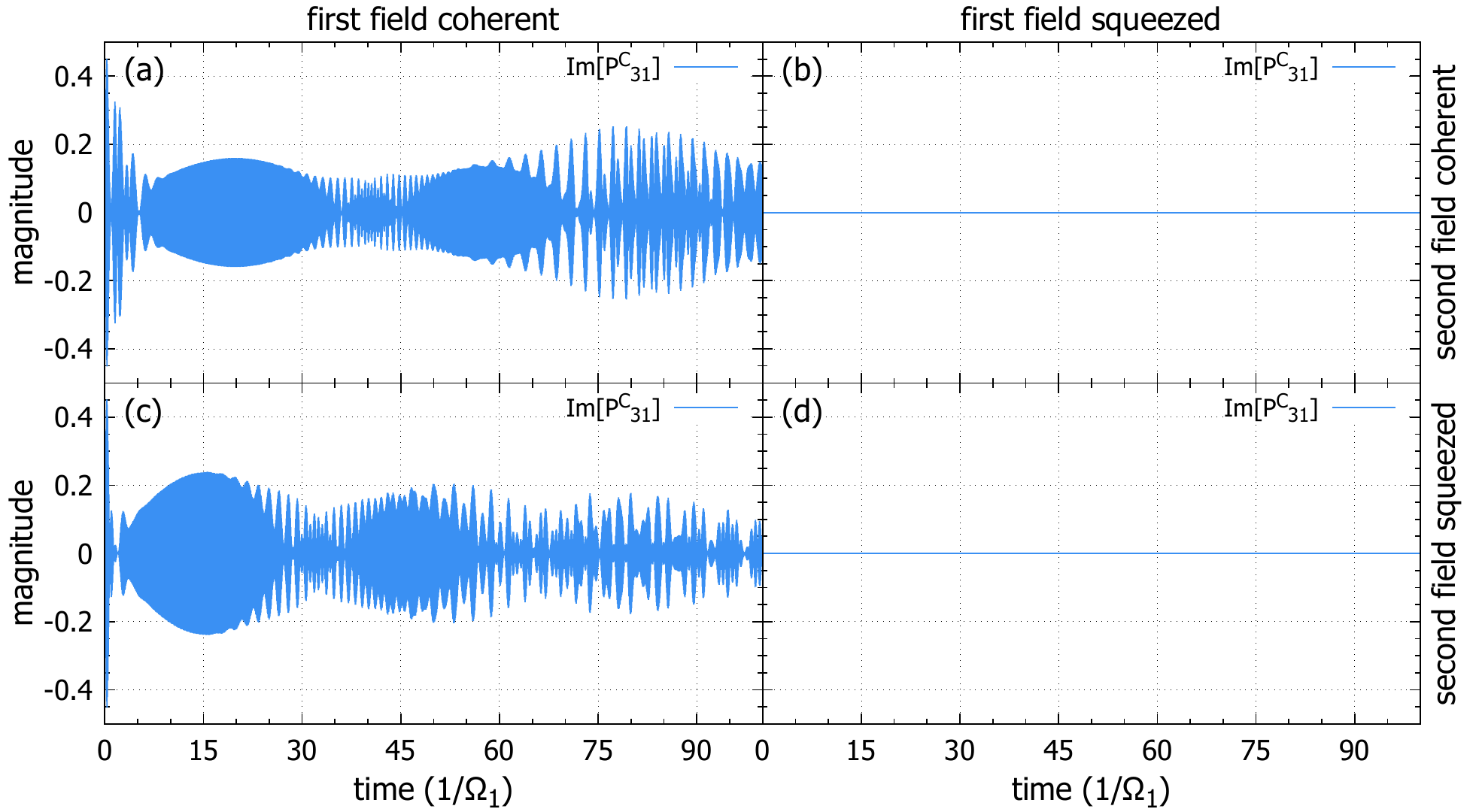}
	\caption{(color online) Dynamics of the classical polarization for the $\ket{3}$-$\ket{1}$ transition without losses and detunings. The four graphs show the dynamics of  $P^C_{31}(t)$  for different initial states of light for the first and the second fields: (a) two coherent states, (b) a squeezed vacuum and a coherent state, (c) a coherent and a squeezed vacuum state, and (d) two squeezed vacuum states. }
	\label{fig:cPol31}
\end{figure*}

\begin{figure*}[ht]
	\centering
	\includegraphics[width = 1.6\columnwidth]{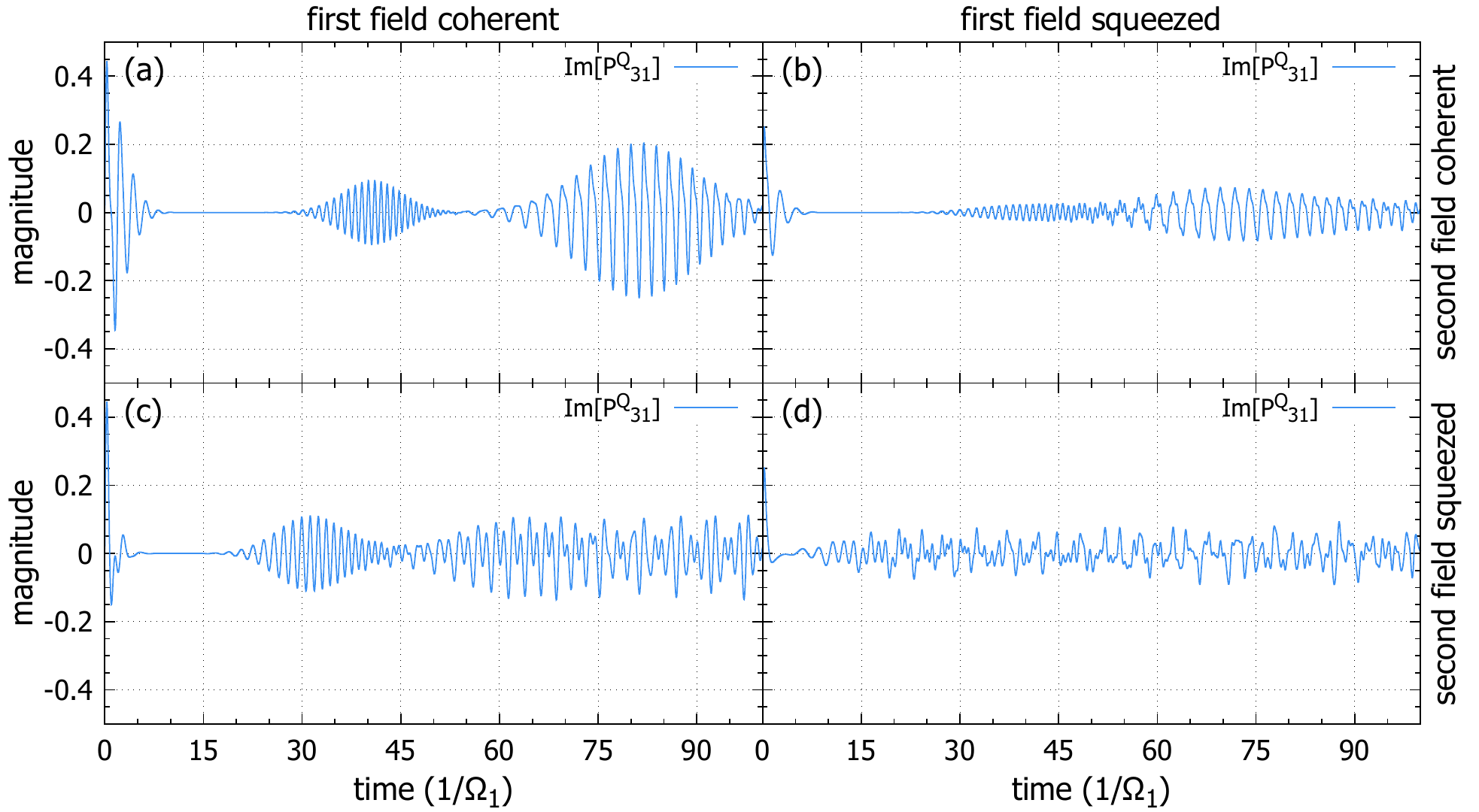}
	\caption{(color online) Dynamics of the quantum polarization for the $\ket{3}$-$\ket{1}$ transition without losses or detunings. The four graphs show the dynamics that arises due to a different initial states of light for the first and the second field: (a) two coherent states, (b) a squeezed vacuum and a coherent state, (c) a coherent and a squeezed vacuum state, and (d) two squeezed vacuum states.}
	\label{fig:qPol31}
\end{figure*}

\subsection{Classical Polarization vs Quantum Polarization}\label{sec:CvQ}

In this section, we investigate the classical polarization $P^C_{31}(t)$ and the quantum polarization $P^Q_{31}(t)$ induced by the first field, which were introduced in Eqs.~(\ref{eq:cp}) and (\ref{eq:qp}), respectively. We start our analysis with the classical polarization $P^C_{31}(t)$ in the absence of any loss mechanisms and for resonant excitation. We consider states of light with a mean photon number $\braket{\hat{n}}=10$. In a semiclassical theory, a resonant excitation leads to a polarization that oscillates with the band gap frequency, i.e., the transition frequency between the states $\ket{1}$ and $\ket{3}$.
This property can be seen in $P^C_{31}(t)$ by extracting the time-dependent exponent from the density matrix elements:
\begin{align}
P^C_{31}(t) = \exp\bigg(-i\frac{E_3 - E_1}{\hbar}t\bigg) \sum_{k=0}^{\infty}\sum_{m=0}^{\infty}p_{\substack{3,k,m \\ 1,k,m}},
\label{eq:decomposition2}
\end{align}
where $\omega_{31} = \frac{E_3 - E_1}{\hbar}$ is the band gap frequency for the $\ket{3}$-$\ket{1}$ transition and $p_{\substack{3,k,m \\ 1,k,m}}$ correspond to a slowly varying envelope of the rapid oscillations with $\omega_{31}$. $\omega_{31}$ has to be larger than all considered detuning values and we assume $\omega_{31} = 100 \Omega_1$  in our numerical simulations.  Fig.~\ref{fig:cPol31} shows the dynamics of $\mathrm{Im}[P^C_{31}(t)]$ for four different combinations of initial states of light. The envelope $p_{\substack{3,k,m \\ 1,k,m}}$ is found to be purely imaginary, the rapid oscillations of the real and imaginary parts of $P^C_{31}(t)$ are caused by the time-dependent exponential factor. This is why, they cannot be distinguished on a long timescale and the real part of $P^C_{31}(t)$ is not shown. 

It is demonstrated by Fig.~\ref{fig:cPol31}, that $P^C_{31}(t)$ is finite when a coherent state excites the $\ket{3}$-$\ket{1}$  transition
but vanishes for a squeezed state. This can be understood from the equations of motion and initial conditions. From the equations of motion (\ref{eq:Motion}) it follows that the dynamics of $\rho_{\substack{3,k,m \\ 1,k,m}}$ is connected with $\rho_{\substack{1,k+1,m \\ 1,k,m}}$, which should be non-zero initially.  The last density matrix elements are proportional to the product of probability amplitudes for measuring $k$ and $k+1$ photons in the first field according to
\begin{align}
\rho_{\substack{1,k+1,m \\ 1,k,m}}(t=0)\propto c_{k+1}c^*_{k}.
\end{align}
Therefore, whether $c_{k+1}c^*_{k}$ is zero or not is the main criterion for whether $P^C_{31}(t)$ is zero or not. Thus, the existence of the classical polarization $P^C_{31}(t)$ is a property of the initial state of the first field. With this criterion, one can conclude that $P^C_{31}(t)$ is non-zero when the first field is a coherent field, while the second field can be arbitrary. For example, the classical polarization $P^C_{31}(t)$ is non-zero when a squeezed state excites the $\ket{3}$-$\ket{2}$ transition while the $\ket{3}$-$\ket{1}$ transition is excited by a coherent state. This statement is also valid for $P^C_{32}(t)$, where the second field needs to be a coherent state. For a finite $P^C_{21}(t)$, both fields need to be coherent, or more precisely none of them can be a squeezed state.

We furthermore note that a classical polarization is not required in order to induce an electronic population in the third level $\ket{3}$, since this can also be generated by two squeezed states, as shown in Section~\ref{sec:pop}. This aligns with the result shown in \cite{PhysRevA.73.013813}, where the generation of an exciton population was demonstrated without a polarization-to-population conversion, by using a thermal state as quantum excitation.

Thus, the classical polarization cannot be a suitable measure for describing the interaction of arbitrary quantum light with matter, and one should introduce a new measure, namely, the quantum polarization $P^Q_{31}(t)$, on which we concentrate below. Fig.~\ref{fig:qPol31} shows $\mathrm{Im}[P^Q_{31}(t)]$ for all four combinations of coherent and squeezed states as initial fields. Since the envelopes $p_{\substack{3,k,m \\ 1,k+1,m}}$ are pure imaginary and the time-dependent exponent is zero for the quantum polarization, the real part is also zero (without the consideration of an optical detuning) and is not shown.  Due to the zero time-dependent exponent, the quantum polarization does not have rapid oscillations as the classical polarization, which oscillates with the band gap frequency $\omega_{31}$. We note that the dynamics of the quantum polarization $P^Q_{31}(t)$ is similar to the population dynamics that was investigated in Section~\ref{sec:pop}. Moreover, the quantum polarization is non-zero for all combinations of the considered initial states of light, which makes $P^Q_{31}(t)$ a suitable measure for analyzing light-matter interaction with quantum light. Collapses and revivals of the quantum polarization are found if at least one of the fields is initially a coherent state.  

The quantum polarization $P^Q_{31}(t)$ can be used to describe the dispersion and absorption of a quantum excitation. We demonstrate this both, numerically by calculating  of the time-averaged real and imaginary part of $P^Q_{31}(t)$ across the detuning range, which is denoted with $P^Q_{31,N}$,  and analytically in the perturbation regime for a very weak probe field. The analytical expression reads:
\begin{align}
P^Q_{31,A}(\Delta_P) = \sum_{k=0}^\infty \sum_{m=0}^\infty \frac{\Omega_1 \Delta_P \sqrt{\frac{k+1}{2}} |c^P_{k+1}|^2|c^C_m|^2}{\Omega^2_2 (\frac{m+1}{2}) - \Delta^2_P - i\Delta_P \tilde{r}_{1,3}},
\label{eq:qpolanalytics}
\end{align}
where $\tilde{r}_{1,3}$ denotes the decay of the non-diagonal matrix elements without additional dephasing as in \cite{Scully}.

\begin{figure}[ht]
	\centering
	\includegraphics[width = 1\columnwidth]{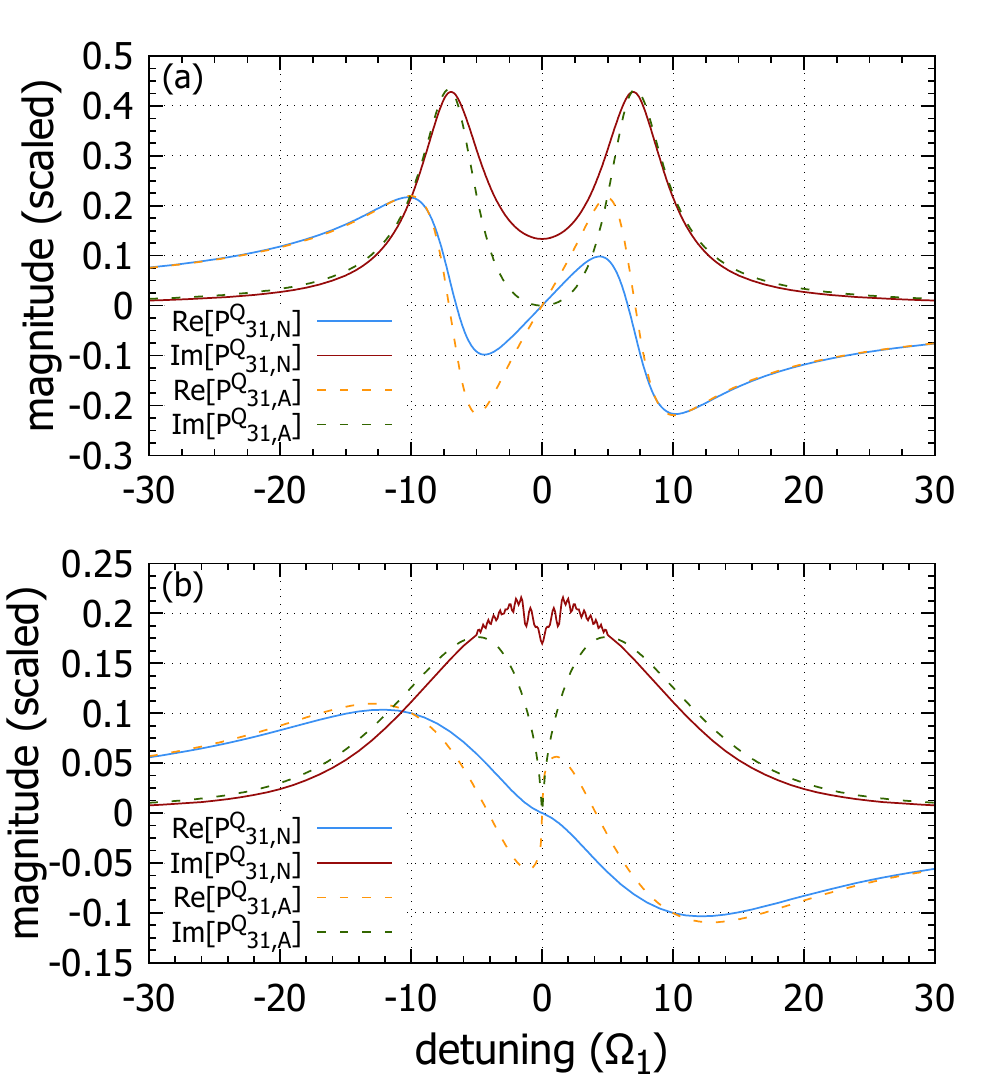}
	\caption{(color online) Time-averaged real and imaginary parts of the numerically calculated quantum polarization $P^Q_{31,N}$ (solid lines) and the real and imaginary parts of the analytical solution $P^Q_{31,A}$ from Eq.~(\ref{eq:qpolanalytics}) (dashed lines). The time-averaging is performed until $t\Omega_1=20$ and no losses are included. The analytical calculation is performed for $\tilde{r}_{1,3}=5\Omega_1$. In (a) two coherent states and in (b) two squeezed vacuum states are considered. The temporal step width is chosen as (a) $\Delta t \Omega_1 = 0.01$, and (b) $\Delta t \Omega_1=0.001$. In all cases $N_P=10$ and $N_C=100$.}
	\label{fig:Fig11}
\end{figure}

Fig.~\ref{fig:Fig11} shows the real and imaginary parts of both, $P^Q_{31,N}$ (solid lines) and $P^Q_{31,A}$ (dashed lines), for two coherent states in (a) and two squeezed vacuum states in (b) with $N_P=10$ and $N_C=100$. The time-averaging is performed until $t\Omega_1=20$, which models a situation in which the signal has decayed after the first collapse of the wave function. We note that Fig.~\ref{fig:Fig11}(a) is similar to the result known from the semiclassical theory \cite{Scully}, however, presents a measure that is exclusively connected with a quantum-optical treatment and applicable to nonclassical states of light in contrast to the classical polarization.

The real parts $\mathrm{Re}[P^Q_{31,N}]$ and $\mathrm{Re}[P^Q_{31,A}]$ are in a good qualitative agreement, since the shapes are similar, but the peaks are more pronounced for the analytical result. We also note that $\mathrm{Re}[P^Q_{31,N}]$ is already converged for the time-averaging up to $t\Omega_1 = 20$. This is not the case for $\mathrm{Im}[P^Q_{31,N}]$, whose value depends on the chosen time intervals and approaches zero for increasingly long time intervals. The choice of the time interval is the reason for the oscillatory behavior of $\mathrm{Im}[P^Q_{31,N}]$ in Fig.~\ref{fig:Fig11}(b). Also note that the fast oscillations are only visible in the range from $\Delta_P=-5\Omega_1$ to $\Delta_P=5\Omega_1$, since it was calculated with a step width of $0.1\Omega_1$, while a step width of $1\Omega_1$ was used otherwise.  In contrary, $\mathrm{Im}[P^Q_{31,A}]$ does not show fast oscillations and is finite since it was derived in the perturbative limit and only represents the linear response. Nevertheless, $\mathrm{Im}[P^Q_{31,A}]$ is in a good qualitative agreement with the spectra shown in Fig.~\ref{fig:EITPop}, where the main difference is that $\mathrm{Im}[P^Q_{31,A}(0)]=0$, which arises from the weak probe field approximation.

Thus, we have found and introduced a novel quantity, namely quantum polarization, which correctly describes the electronic response after the impact of any initial quantum field states and presents a different approach for the demonstration of EIT.

\subsection{Entanglement between Fields with Losses}

\begin{figure}[ht]
	\centering
	\includegraphics[width = 0.8\columnwidth]{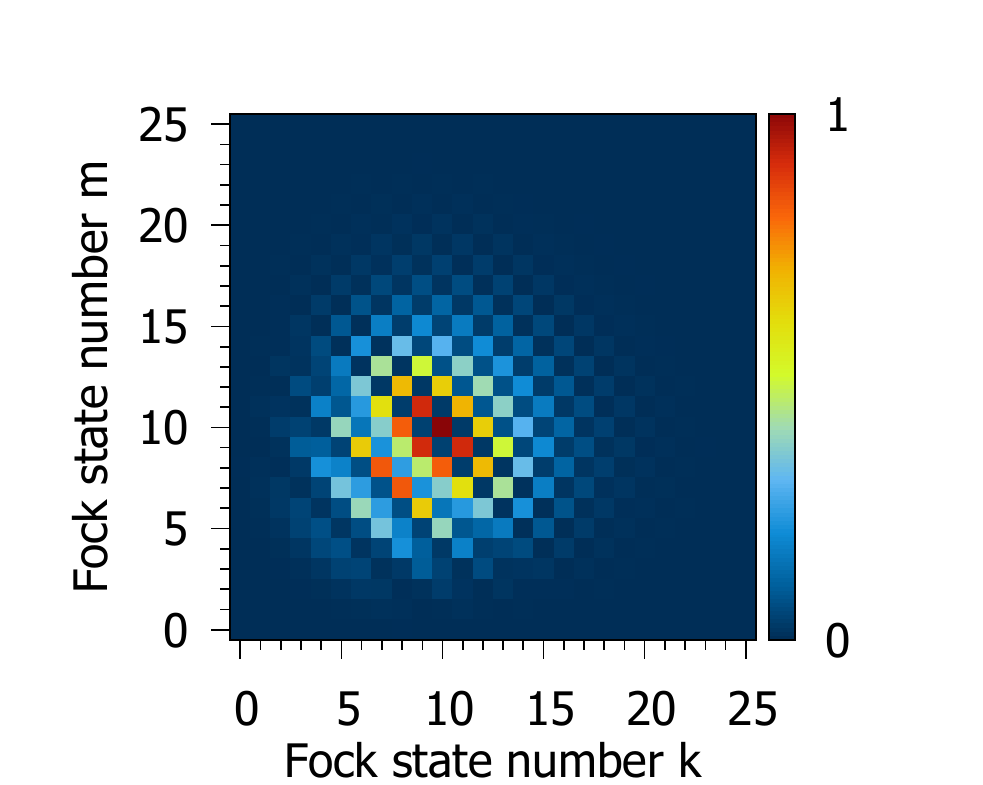}
	\caption{(color online) Bipartite photon number distribution $W_{km}$ at $t\Omega_1 = 23.21$ for the lossless case. The magnitude is normalized to a maximum value of $1$. Two coherent states with a mean photon number of $\braket{\hat{n}}=10$ are considered.}
	\label{fig:Fig12}
\end{figure}

In Ref.~\cite{popolitova2019}, it was demonstrated that the bipartite photon number distribution for the third electronic level $W_{km}=\rho_{\substack{3,k,m \\ 3,k,m}}$ is a suitable representation for the correlation between photons of the two quantum fields, i.e., their entanglement. Fig.~\ref{fig:Fig12} demonstrates $W_{km}$ without losses for two coherent states with a mean photon number of $\braket{\hat{n}}=10$ at $t\Omega_1=23.21$, which is during the first collapse of the population dynamics. The non-Gaussian shape is a clear demonstration of entanglement between the quantum fields that arises due to the light-matter interaction.

\begin{figure}[ht]
	\centering
	\includegraphics[width = 0.95\columnwidth]{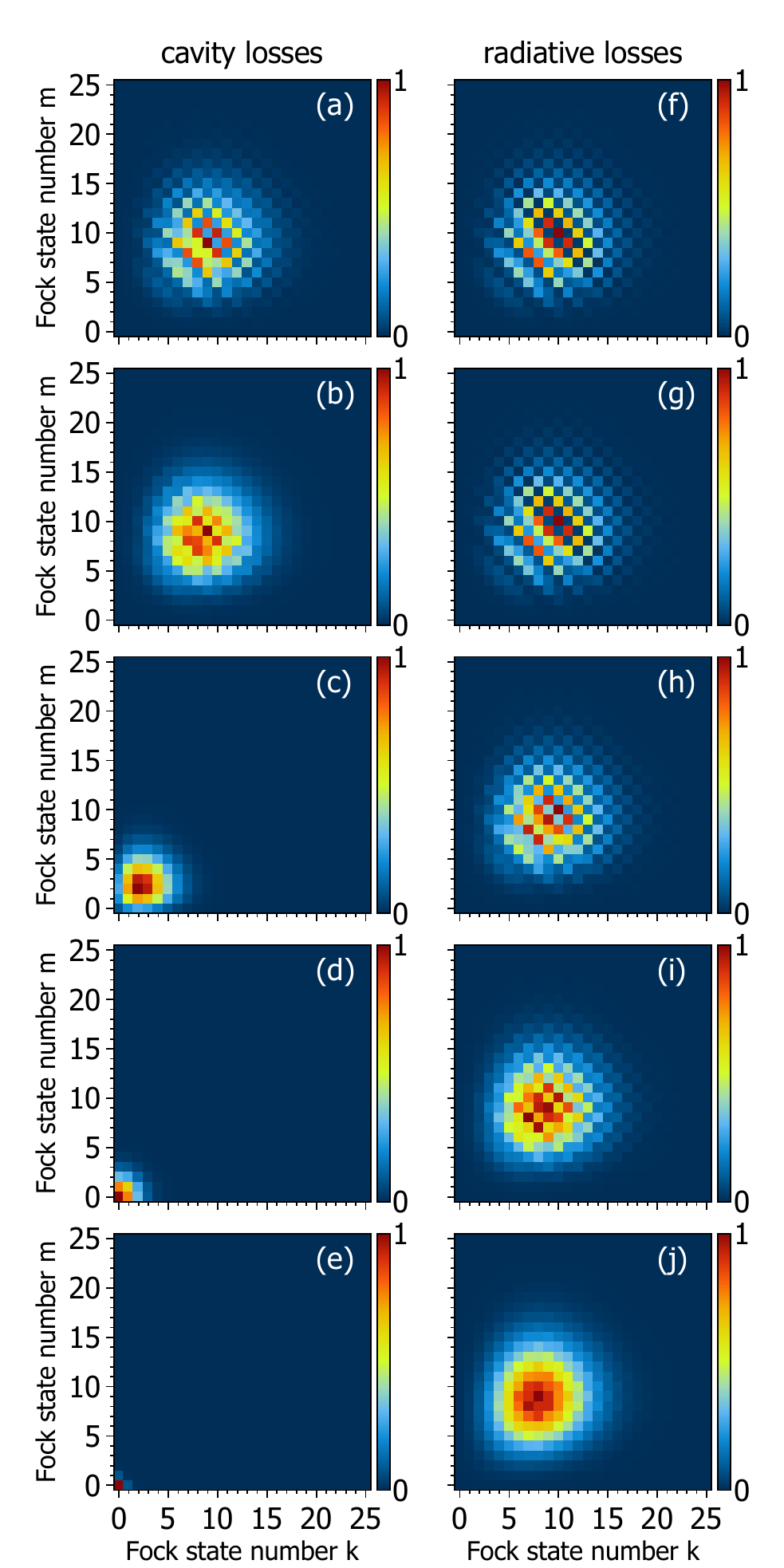}
	\caption{(color online) Bipartite photon number distribution $W_{km}$ at $t\Omega_1 = 23.21$ with losses applied. For (a)-(e) only cavity losses $\kappa$ are applied, while for (f)-(j) only radiative losses $r_{1,3}=r_{2,3}=r$ and $5r_{1,2}=r$ are included. The loss parameters $\kappa$ and $r$ are chosen as (a) and (f) $0.001\Omega_1$, (b) and (g) $0.003\Omega_1$, (c) and (h) $0.05\Omega_1$, (d) and (i) $0.1\Omega_1$, and (e) and (j) $0.2\Omega_1$. The magnitudes are individually normalized to a maximum value of $1$. Two coherent states with mean photon numbers of $\braket{\hat{n}}=10$ are considered.}
	\label{fig:Fig13}
\end{figure}

In the following, we examine the behavior of $W_{km}$ in the presence of radiative and cavity losses. Figs.~\ref{fig:Fig13}(a)-(e) show the same scenario as in Fig.~\ref{fig:Fig12}, but with cavity losses applied with a value of $\kappa_1=0.001\Omega_1$, $\kappa_2=0.003\Omega_1$, $\kappa_3=0.05\Omega_1$, $\kappa_4=0.1\Omega_1$, and $\kappa_5=0.2\Omega_1$, respectively. We see that with ascending cavity losses, the bipartite photon number distribution not only strives towards a Gaussian shape, but also the mean photon number decreases, until the bipartite photon number distribution eventually approaches the vacuum state, see Fig.~\ref{fig:Fig13}(e). This corresponds to the loss of all photons in the system and the destruction of entanglement between the two quantum fields, since a Gaussian distribution is a factorized state.

This analysis is repeated for the case of radiative losses. Here, we always choose $r_{1,3}=r_{2,3}=r$ and $5r_{1,2}=r$, which fits to the parameter sets chosen before. Figs.~\ref{fig:Fig13}(f)-(j) demonstrate how the result of Fig.~\ref{fig:Fig12} is modified by radiative losses. We assign the same value to the loss parameters $r$, as we did for $\kappa$, so that a row in Fig.~\ref{fig:Fig13} correspond to the same loss parameter for cavity and radiative losses, respectively. One can see that in comparison to cavity losses, a higher radiative loss parameter $r$ is required in order to destroy the non-Gaussian shape. The cavity losses reduce entanglement more efficiently than radiative losses.

This conclusion can be understood from the physical mechanisms described by these losses. While cavity losses explicitly destroy photons and therefore directly influence a bipartite photon number distribution, radiative losses destroy the excitation of a transition. As a result, the electronic system has a transition from the excited to the ground state without a change in the number of photons. This destruction leads to a redistribution of the fields and to the loss of entanglement. Finally, we obtain a completely uncorrelated state, where distributions of the first and the second fields are not correlated with each other, see Fig.~\ref{fig:Fig13}(j). Therefore, this result is consistent with the physical interpretation of the investigated loss mechanisms.

\section{CONCLUSION}

In this work, using a Jaynes-Cummings-type model, we investigated the interaction between a 3LS and quantum light. We found that CPT can be achieved by applying radiative losses, without the requirement of an initial coherent superposition of electronic states, which simplifies the experimental realization
that should be possible in atomic systems and semiconductor nanostructures.

By applying higher-order losses to the system, we  demonstrated a redistribution of the photon statistics between the two quantum fields, which in an ideal case can lead to a one-by-one transfer of the statistics. In a more realistic description, key features of the photon statistics are transferred. 

Moreover, we demonstrated the EIT effect in the quantum-optical regime by identifying the time-averaged population as a measure for absorption. The EIT spectrum obtained with coherent states is similar to the well-known result from a semiclassical description but is not its direct representation. In the quantum regime, properties of the probe field directly influence EIT: the EIT effect can be improved using a squeezed probe field, when other parameters are the same. This especially plays a role for the center of the spectrum, where a small absorption is desired. The energetic position of the EIT peaks can be understood from the quasienergy states and can be estimated analytically.

Furthermore, we introduce and analyze a novel quantity, the quantum polarization, which describes the response of mater on quantum fields and therefore contains an information about the quantum field properties. We compared the classical polarization with the quantum polarization and show that the classical polarization vanishes for certain quantum states of light in contrast to the quantum polarization. This makes the quantum polarization a valuable quantity in the context of a quantum-optical description, e.g. allowing to calculate the absorption and dispersion of a quantum excitation. In addition, a qualitative approach for the visualization of quantum correlations between fields was used to demonstrate the impact of losses. It was shown that cavity losses destroy correlations faster than radiative losses, which originates from the mechanism that they are describing.

Our results open new possibilities for characterizing the response of matter interacting with arbitrary quantum light, as well as for manipulating electronic state populations and field statistics using losses which can be used to store quantum memory and transfer quantum information.

\section{ACKNOWLEDGMENTS}

The joint grant by the Deutsche Forschungsgemeinschaft (DFG) and the Russian Science Foundation (RSF) (projects SH 1228/2-1, ME 1916/7-1, No.19-42-04105) is gratefully acknowledged.

\end{document}